\documentclass[fleqn,10pt]{wlscirep}
\usepackage[utf8]{inputenc}
\usepackage[T1]{fontenc}
\usepackage{glossaries}

\def\RP{La$_{4}$Ni$_{3}$O$_{10}$} 
\def\RRP{La$_{4}$Ni$_{3}$O$_{8}$} 

\newacronym{CO}{CO}{charge order}
\newacronym{JDOS}{JDOS}{joint density of states}
\newacronym{BS}{BS}{bond-stretching}
\newacronym{RIXS}{RIXS}{resonant inelastic x-ray scattering}
\newacronym{REXS}{REXS}{resonant elastic x-ray scattering}
\newacronym{XAS}{XAS}{x-ray absorption spectroscopy}
\newacronym{XLD}{XLD}{x-ray linear dichroism}
\newacronym{TEY}{TEY}{total electron yield}
\newacronym{TFY}{TFY}{total fluorescence yield}
\newacronym{EELS}{EELS}{electron energy loss spectroscopy}
\newacronym{EPC}{EPC}{electron-phonon coupling}
\newacronym{CDW}{CDW}{charge density wave}
\newacronym{SDW}{SDW}{spin density wave}
\newacronym{FWHM}{FWHM}{full-width at half-maximum}
\newacronym{INS}{INS}{inelastic neutron scattering}
\newacronym{DFT}{DFT}{density functional theory}
\newacronym{GGA}{GGA}{generalized gradient approximation}
\newacronym{UHB}{UHB}{upper Hubbard band}
\newacronym{ZSA}{ZSA}{Zaanen-Sawatzky-Allen}
\newacronym{ZRS}{ZRS}{Zhang-Rice singlet}
\newacronym{ED}{ED}{exact diagonalization}
\newacronym{CEF}{CEF}{crystal electric field}
\newacronym{2D}{2D}{two-dimensional}
\newacronym{TM}{TM}{transition-metal}
\newacronym{LDA}{LDA}{local density approximation}
\newacronym{DMFT}{DMFT}{dynamical mean field theory}

\title{Data descriptor:\\
Resonant inelastic x-ray scattering data for Ruddlesden-Popper and reduced Ruddlesden-Popper nickelates}

\author[1,2,*]{G.~Fabbris}
\author[1, 3]{D.~Meyers}
\author[1]{Y.~Shen}
\author[4]{V.~Bisogni}

\author[5,6]{J.~Zhang}
\author[5]{J.~F.~Mitchell}

\author[5]{M.~R.~Norman}
\author[7,8]{S.~Johnston}

\author[9, +]{J.~Feng}
\author[9,10]{G.~S.~Chiuzb\u{a}ian}
\author[10]{A.~Nicolaou}
\author[10]{N.~Jaouen}
\author[1,*]{M.~P.~M.~Dean}

\affil[1]{Condensed Matter Physics and Materials Science Department, Brookhaven National Laboratory, Upton, New York 11973, USA}
\affil[2]{Advanced Photon Source, Argonne National Laboratory, Lemont, Illinois 60439, USA}
\affil[3]{Department of Physics, Oklahoma State University, Stillwater, Oklahoma 74078, USA}
\affil[4]{National Synchrotron Light Source II, Brookhaven National Laboratory, Upton, New York 11973, USA}
\affil[5]{Materials Science Division, Argonne National Laboratory, Lemont, Illinois 60439, USA}
\affil[6]{Institute of Crystal Materials, Shandong University, Jinan, Shandong 250100, China}
\affil[7]{Department of Physics and Astronomy, The University of Tennessee, Knoxville, Tennessee 37966, USA}
\affil[8]{Institute of Advanced Materials and Manufacturing, The University of Tennessee, Knoxville, Tennessee 37996, USA}
\affil[9]{Sorbonne Universit\'{e}, CNRS, Laboratoire de Chimie
Physique-Mati\`{e}re et Rayonnement, 75005 Paris, France}
\affil[10]{Synchrotron SOLEIL, L’Orme des Merisiers, Saint-Aubin, BP 48, 91192 Gif-sur-Yvette, France}

\affil[*]{corresponding authors: G. Fabbris (gfabbris@anl.gov) and M. P. M. Dean (mdean@bnl.gov)}

\affil[+]{Institute of Advanced Science Facilities, Shenzhen 518000, China}

\begin{abstract}
Ruddlesden-Popper and reduced Ruddlesden-Popper nickelates are intriguing candidates for mimicking the properties of high-temperature superconducting cuprates. The degree of similarity between these nickelates and cuprates has been the subject of considerable debate. \Gls*{RIXS} has played an important role in exploring their electronic and magnetic excitations, but these efforts have been stymied by inconsistencies between different samples and the lack of publicly available data for detailed comparison. To address this issue, we present open \gls*{RIXS} data on \RP{} and \RRP{}. 
\end{abstract}
\begin{document}

\flushbottom
\maketitle
\thispagestyle{empty}

\section*{Background \& Summary}

\begin{figure}[ht]
\centering
\includegraphics[width=0.9\linewidth]{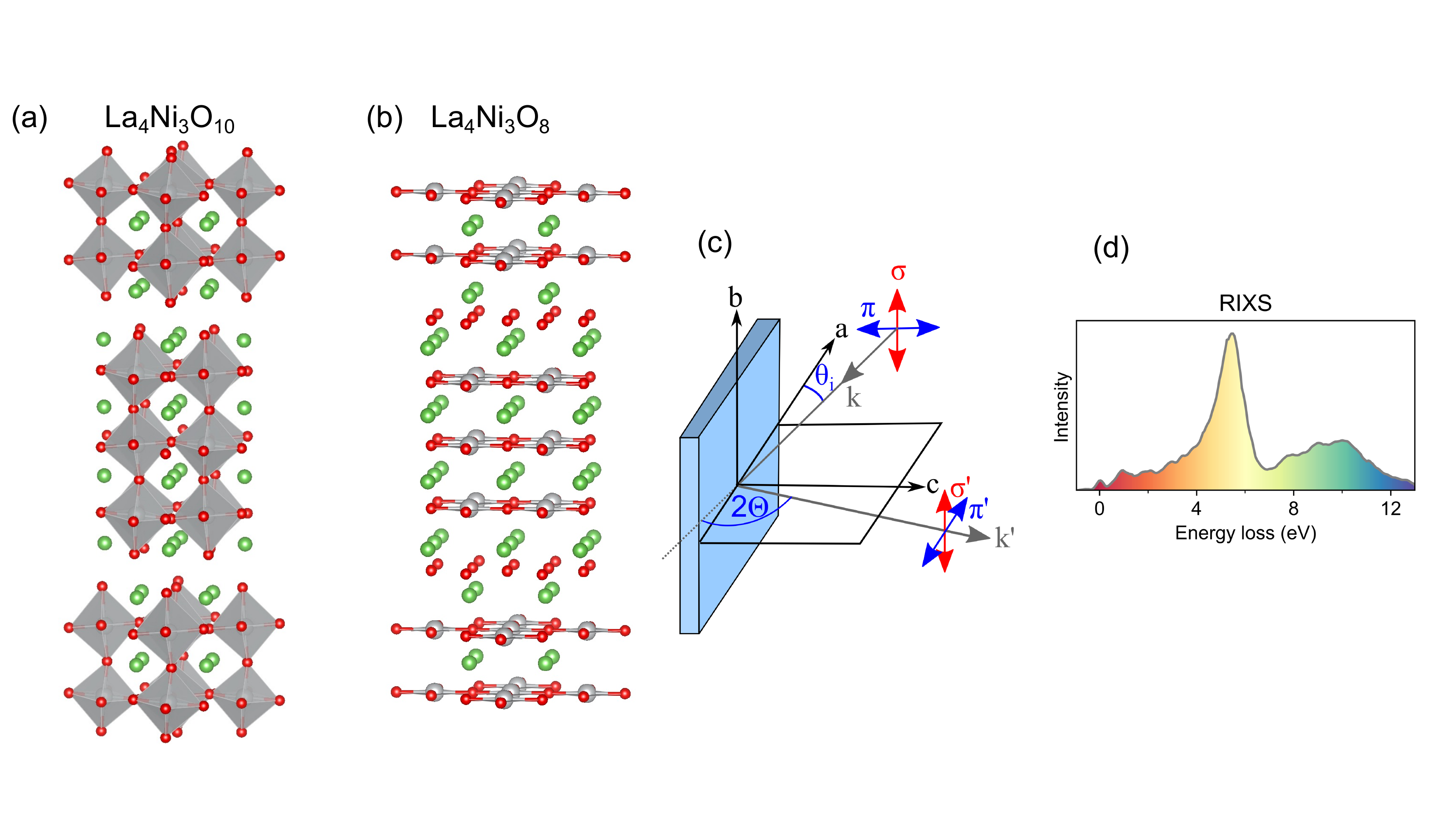}
\caption{Summary of study. (\textbf{a}),(\textbf{b}) Crystal structure of (\textbf{a}) \RP{} and (\textbf{b}) \RRP{}. La, Ni, and O atoms are depicted by green, gray, and red spheres, respectively. (\textbf{c}) The experimental geometry showing the incident x-ray angle $\theta_i$, the scattering angle $2\Theta$, and incident x-ray polarization $\pi$ and $\sigma$. (\textbf{d}) An example \gls*{RIXS} spectrum, which encodes the electronic structure of these materials. }
\label{fig:intro}
\end{figure}

Generating analogs of the copper-based high-temperature superconductors based on different transition metal ions has been a major goal of materials research since the 1980s, as this opens routes to improving our understanding of unconventional superconductivity and realizing materials with improved properties \cite{Anisimov1999electronic, Norman2016materials, Adler2018correlated}. Since nickel lies immediately to the left of copper in the periodic table, it represents an obvious target element for realizing new superconductors. However, despite extensive work, no superconductivity has been found in nominally Ni $3d^8$ or $3d^7$ compounds \cite{Chaloupka2008orbital, Hansmann2009turninng, Disa2015orbital}. One route to realizing the desired $3d^9$ state found in copper-based superconductors in a nickel oxide has been achieved by preparing members of the Ruddlesden-Popper phase materials with formula $R_{n+1}$Ni$_n$O$_{3n+1}$ (where $R$ is a rare earth element) and reducing them through removal of their apical oxygen atoms, forming $R_{n+1}$Ni$_n$O$_{2n+2}$ materials [see Fig.~\ref{fig:intro}(a),(b)]. Indeed, this approach has been validated by the discovery of superconductivity in infinite layer $n=\infty$ \textit{R}$_{1-x}$Sr$_x$NiO$_2$, and subsequently $n=5$ Nd$_6$Ni$_5$O$_{12}$, generating intense scientific interest \cite{Li2019superconductivity, Osada2020superconducting, Pan2021super, Zeng2022Superconductivity}. 

Resonant x-ray spectroscopy offers several advantages for probing low valence nickelates \cite{Hepting2021soft}. It can probe small sample volumes, which is useful since these materials are challenging to prepare as large single crystals. Indeed, many low valence nickelate samples are only available in thin film form. Another unique feature is the ability of \gls*{RIXS} to selectively study different atomic species through the resonant process and its access to magnetic and orbital excitations, which are optically forbidden  \cite{Dean2015insights}. Efforts to understand the detailed electronic properties of Ruddlesden-Popper and reduced Ruddlesden-Popper nickelates have consequently relied heavily on resonant x-ray spectroscopy \cite{Fabbris2017doping, Zhang2017large, Hepting2020electronic, Lin2021strong, Lu2021magnetic, Higashi2021core, Rossi2020orbital, Shen2022role, Norman2023orbital, Shen2023orbital, Hepting2021soft, Wang2019EDRIXS}.

However, much of this debate has been complicated by two major difficulties. The first is inconsistent spectra from nominally equivalent or very similar samples, an issue raised specifically in Refs.~\cite{Higashi2021core, Krieger2021Charge}. The second is the lack of open data, which precludes detailed direct comparison between samples. Herein, we provide \gls*{RIXS} data from bulk crystals of \RP{} ($n=3$) and \RRP{} ($n=3$, after reduction). In addition, little publicly accessible advice is currently available for comparing \gls*{RIXS} datasets, so we take this opportunity to outline considerations in comparing \gls*{RIXS} data taken under different experimental conditions.

\section*{Methods}

The parent Ruddlesden-Popper \RP{} was synthesized using the high pressure optical floating zone method at Argonne National Laboratory. During growth, 0.1~l/min of oxygen gas was flowed through the reaction chamber at 20~bar pressure and the feed rods were advanced at 4~mm/h over the 30 hour growth time. To improve homogeneity, the feed and seed rods were counter-rotated at 30~r.p.m. throughout this process. A similar procedure was used in Ref.~\cite{Zhang2016stacked}. The $c$-axis surface used in the measurement was mirror polished in water using diamond lapping paper. \RRP{} single crystals were prepared by cleaving small crystals from the parent \RP{} crystals and reducing them by heating in a flowing 4\% H$_2$/Ar gas mixture at 350$^{\circ}$C for five days. This procedure was used successfully in several prior works \cite{Zhang2016stacked, Zhang2017large, Lin2021strong, Shen2022role, Shen2023orbital}.

Ni $L_{3,2}$ \gls*{RIXS} as well as O $K$ and Ni $L_{3,2}$ \gls*{XAS} measurements were performed at the AERHA instrument of the SEXTANTS beamline at the SOLEIL synchrotron. The spectrometer operates by dispersing the x-ray photons as a function of their energy onto a two-dimensional detector via the detailed optical scheme described in \cite{Chiuzbaian2014}. A sketch of the experimental geometry is displayed in Fig.~\ref{fig:intro}(c). All measurements were performed with the $a$ and $c$ sample axes in the horizontal scattering plane. At the Ni $L$ edges, data were taken at $\theta = 15^\circ$ degrees ($\theta$ is the angle between the incoming x-ray and the sample surface), the linear dichroism is obtained by switching the incident x-ray polarization between $\pi$ and $\sigma$. For the O $K$ edge, a fixed $\pi$ x-ray polarization was used, the dichroism is obtained by measuring at $\theta = 15^\circ$ and $90^\circ$. \gls*{RIXS} was measured at $2\Theta = 95^{\circ}$, with an overall \gls*{FWHM} energy resolution of $262 \pm 9$ meV, chosen in order to balance throughput and resolution. As is standard in modern \gls*{RIXS} experiments, the two-dimensional data are binned in the isoenergetic direction to form spectra and the pixel to energy loss conversion is performed by measuring the elastic line of the spectrometer while changing the beamline energy. The data describing this calibration are provided in the Technical Validation section.

\section*{Data Records}

\begin{figure}[ht]
\centering
\includegraphics[width=\linewidth]{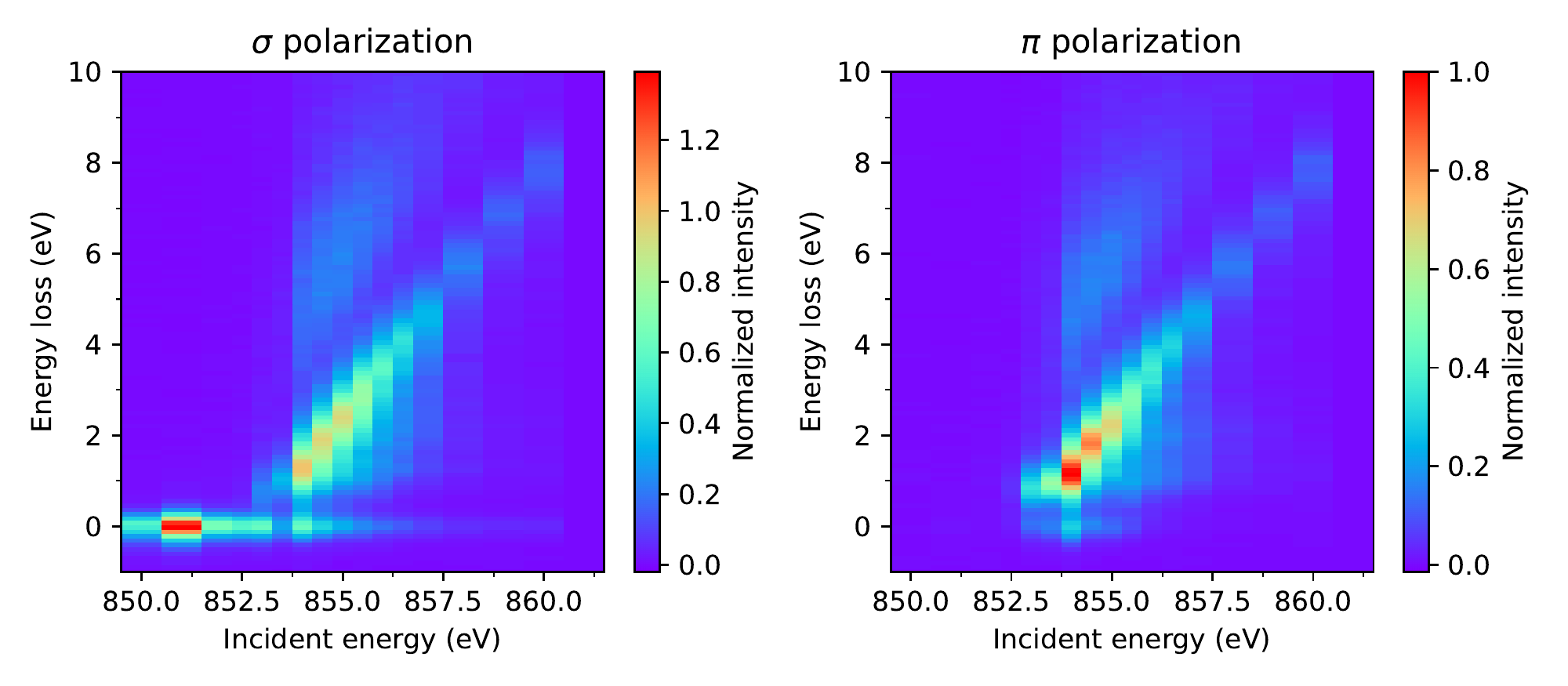}
\caption{Ni $L_3$-edge \gls*{RIXS} energy map for \RP{} measured at 25~K with an incident angle of $\theta_i=15^\circ$ and scattering angle of $2\Theta=95^\circ$. The left and right panels show $\sigma$ and $\pi$ incident x-ray polarization, respectively.}
\label{fig:RP}
\end{figure}

\begin{figure}[ht]
\centering
\includegraphics[width=\linewidth]{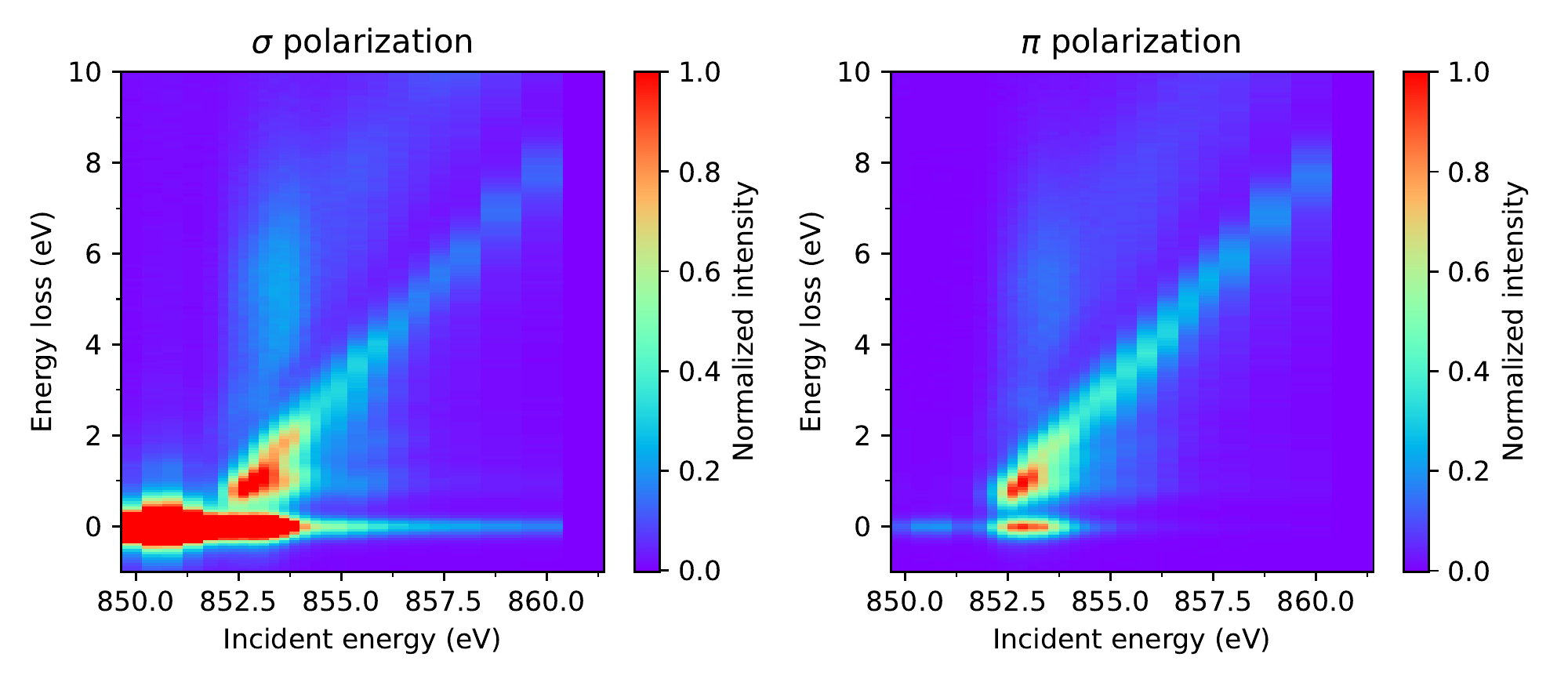}
\caption{Ni $L_3$-edge \gls*{RIXS} energy map for \RRP{} measured at 25~K with an incident angle of $\theta_i=15^\circ$ and scattering angle of $2\Theta=95^\circ$. The left and right panels show $\sigma$ and $\pi$ incident x-ray polarization, respectively.}
\label{fig:RRP}
\end{figure}

\begin{figure}[ht]
\centering
\includegraphics[width=\linewidth]{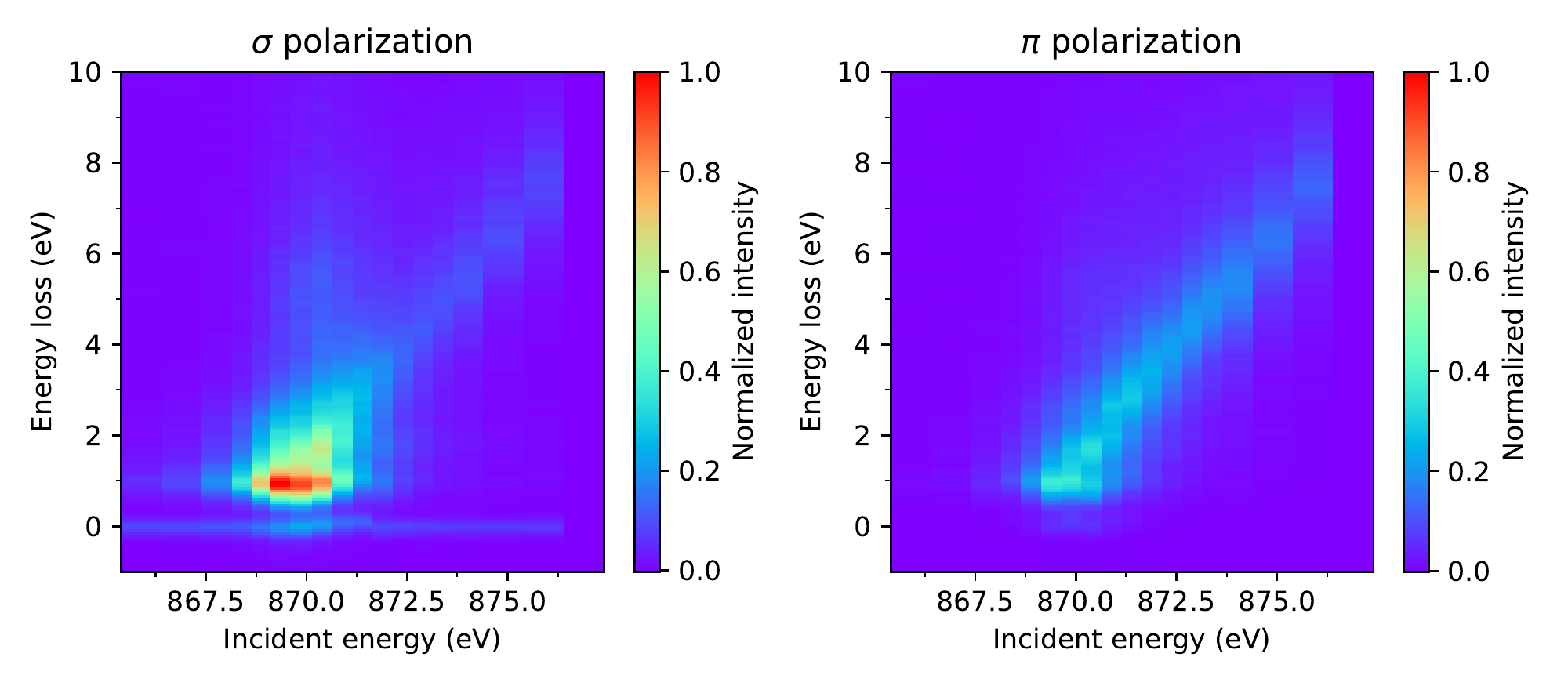}
\caption{Ni $L_2$-edge \gls*{RIXS} energy map for \RRP{} measured at 25~K with an incident angle of $\theta_i=15^\circ$ and scattering angle of $2\Theta=95^\circ$. The left and right panels show $\sigma$ and $\pi$ incident x-ray polarization, respectively.}
\label{fig:RRP_L2}
\end{figure}

This Data Descriptor is based on data deposited in the Zenodo general-purpose open repository \cite{Zenodo}. \gls*{RIXS} data are provided as text files with filename of the form \texttt{La4Ni3O10\_pi\_850.0eV.dat}, which specify the material, incident x-ray polarization, and incident energy. This information as well as the sample temperature are provided in the file header. \gls*{XAS} data are also provides as text files with a similar nomenclature, but which includes the absorption edge measured, for instance \texttt{La4Ni3O8\_sigma\_OKedgeXAS.dat}. These data are also plotted in Figs.~\ref{fig:RP}, \ref{fig:RRP}, and \ref{fig:RRP_L2}. The \gls*{XAS} data displayed in Fig.~\ref{fig:xas} were taken in \gls*{TFY}, but files also contain data taken in \gls*{TEY} concomitantly. Finally, the data shown in Fig.~\ref{fig:elastic}(a)\&(b) are stored in files \texttt{elastic\_850eV.dat} and \texttt{pixel\_energy\_dispersion.dat}, respectively. The repository illustrates the data content using plotting scripts based on standard python stack of numpy, matplotlib, scipy, and pandas \cite{harris2020array, Hunter2007matplotlib, mckinney2010data}. The jupyter-repo2docker is used to define the computational environment and making the code executable on services such as \href{https://mybinder.org}{mybinder.org}.

\section*{Technical Validation}

\begin{figure}[ht]
\centering
\includegraphics[width=\linewidth]{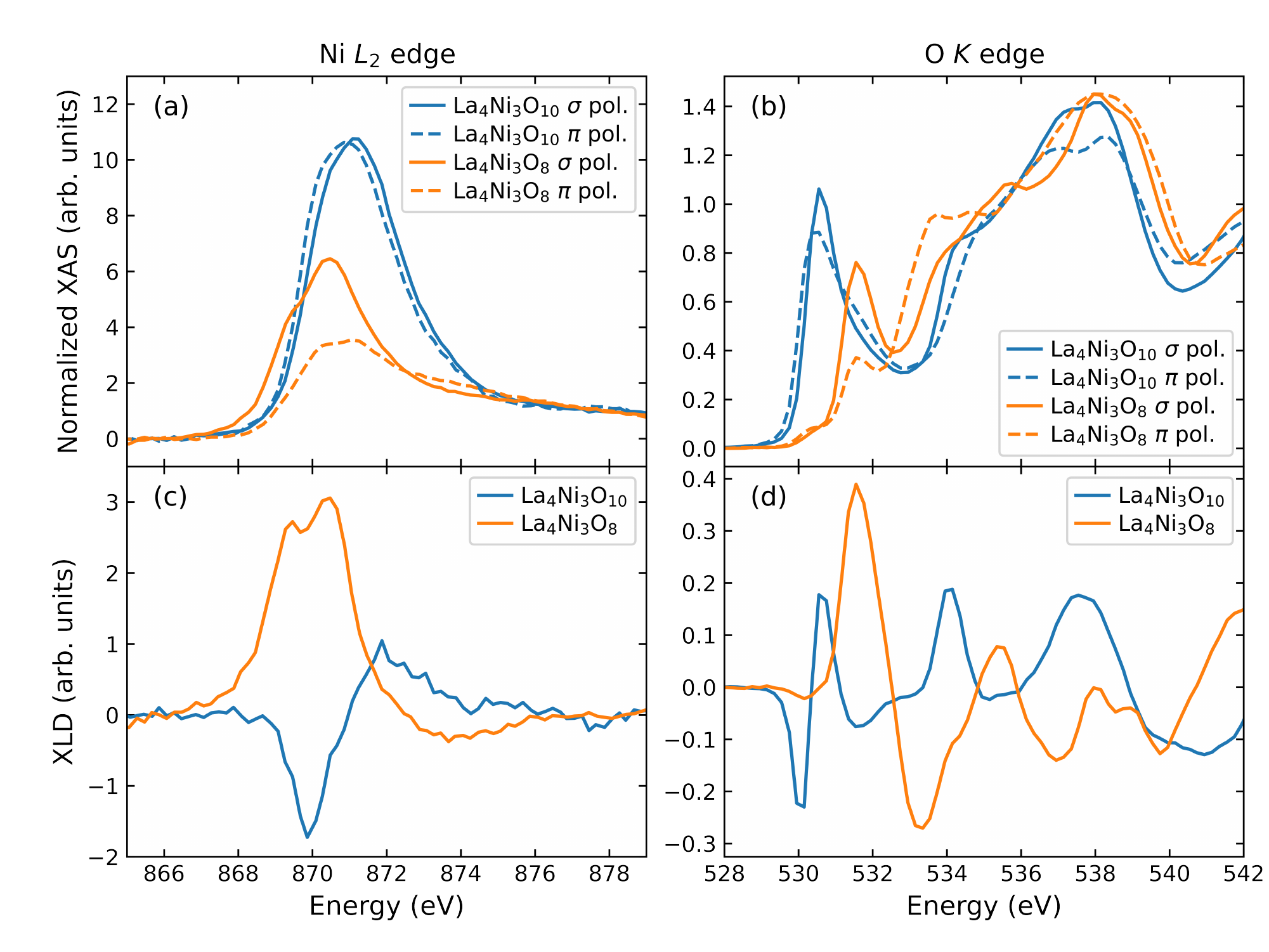}
\caption{\gls*{XAS} measurements of \RP{} and \RRP{} taken at 300 K and $\theta_i = 15^{\circ}$. The Ni $L_2$ edge \gls*{XAS} and \gls*{XLD} are displayed in panels (\textbf{a})\&(\textbf{c}), respectively. Panels (\textbf{b})\&(\textbf{d}) display the O $K$ edge \gls*{XAS} and \gls*{XLD}, respectively.}
\label{fig:xas}
\end{figure}

\begin{figure}[ht]
\centering
\includegraphics[width=\linewidth]{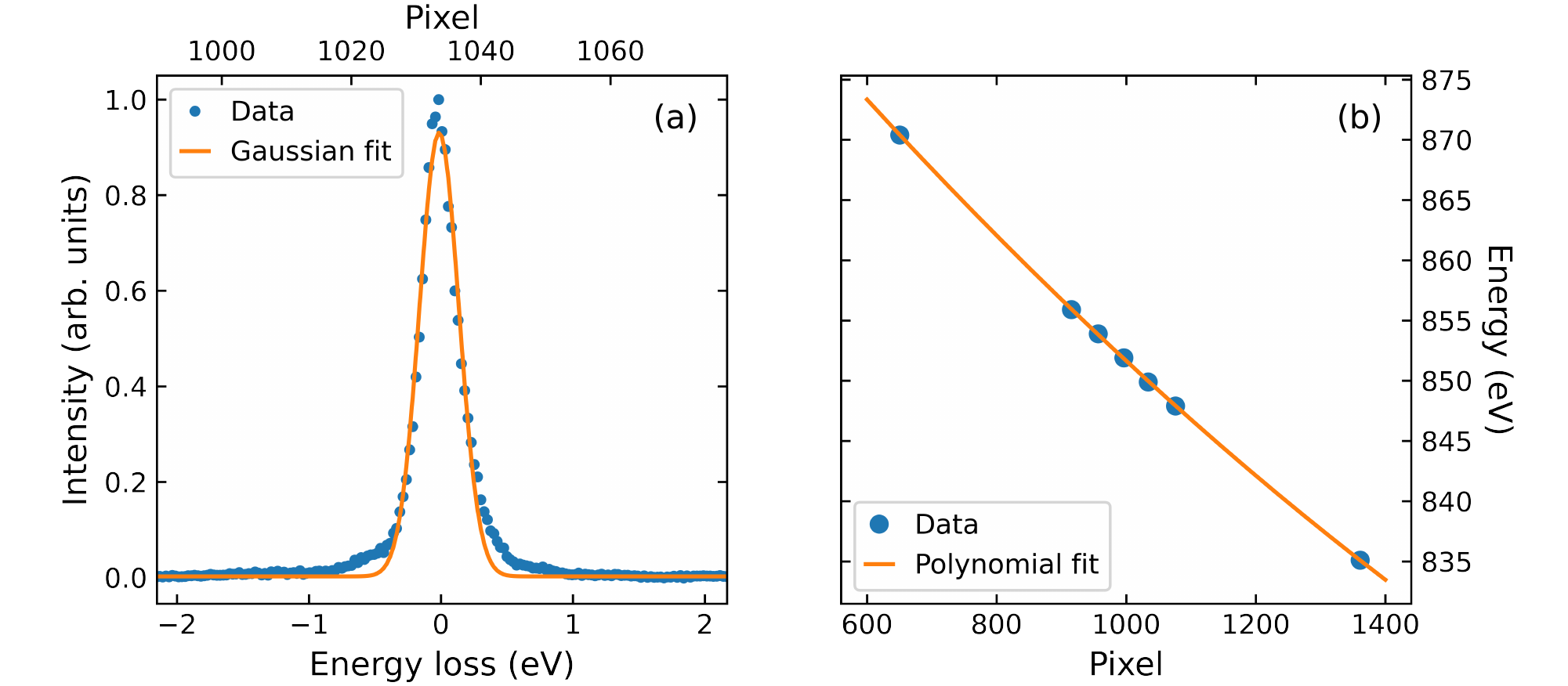}
\caption{\gls*{RIXS} instrument parameters. (\textbf{a}) Elastic line measurement; a gaussian fit yields \gls*{FWHM} = $262 \pm 9$ meV. (\textbf{b}) Dispersion of the elastic line through the CCD detector, the calibration follows a second order polynomial ($y = a + bx + cx^2$) with $a = 912.7$ eV, $b = -7.24 \times 10^{-2}$ eV/pixel, and $c = 1.13 \times 10^{-5}$ eV/pixel$^2$.}
\label{fig:elastic}
\end{figure}

Figure \ref{fig:xas} displays the Ni $L_2$ and O $K$ edge \gls*{XAS} and \gls*{XLD} of \RP{} and \RRP{}. The larger \gls*{XAS} white line area seen in \RP{} and the larger \gls*{XLD} in \RRP{} are consistent with previous results \cite{Zhang2017large}. A similarly large \gls*{XLD} in \RRP{} is observed at the O $K$-edge [Fig.~\ref{fig:xas}(d)]. These results confirm the quality of the samples used here. The energy resolution of the \gls*{RIXS} measurement was determined by fitting a Gaussian peak to a spectrum collected at an energy away from the resonance [Fig.~\ref{fig:elastic}(a)]. The data collected by the CCD area detector was converted to energy space using the calibration shown in Fig.~\ref{fig:elastic}(b). Since the La $M_4$-and Ni $L_3$-edges overlap in energy, we also provide data at the Ni $L_2$-edge in Fig.~\ref{fig:RRP_L2}.

\section*{Usage Notes}

Several technical issues make it quite difficult to measure \Gls*{RIXS} in absolute units. We therefore present data in arbitrary units. We suggest comparing the preset spectra to other works after dividing both spectra by the integrated spectral intensity in the 0.5-15 eV energy range. It is better to exclude the elastic line from such comparisons, as the quasielastic intensity is particularly sensitive to extrinsic issues such as sample surface flatness. 

To compare the data presented here with other data, care should be taken to account for possible differences in energy resolution between the different datasets. We suggest comparing data by convolving the higher-resolution spectrum with a lineshape that gives both spectra the same effective resolution.

\section*{Code availability}
The data reported here were generated via synchrotron experiments and did not require any processing of datasets beyond trivial binning of the two-dimensional data into one-dimensional spectra and calibration of the energy loss.

\section*{Acknowledgements}
Work at Brookhaven and the University of Tennessee (RIXS measurements and the interpretation) was supported by the U.S.\ Department of Energy, Office of Science, Office of Basic Energy Sciences, under Award Number DE-SC0022311. Work at Argonne was supported by the U.S.\ DOE, Office of Science, Basic Energy Sciences, Materials Science and Engineering Division (nickelate sample synthesis and \gls*{RIXS} theory). We acknowledge SOLEIL for provision of synchrotron radiation facilities. This research used resources of the Advanced Photon Source, a U.S. Department of Energy (DOE) Office of Science user facility at Argonne National Laboratory and is based on research supported by the U.S. DOE Office of Science-Basic Energy Sciences, under Contract No. DE-AC02-06CH11357. This research used resources at the SIX beamline of the National Synchrotron Light Source II, a U.S.\ DOE Office of Science User Facility operated for the DOE Office of Science by Brookhaven National Laboratory under Contract No.~DE-SC0012704. G.S.C.\ acknowledges the support of the Agence Nationale de la Recherche (ANR), under Grant No.\ ANR-05-NANO-074 (HR-RXRS).

\section*{Author contributions statement}
The project was conceived and supervised by G.F.\ \& M.P.M.D. G.F., D.M., Y.S., V.B., J. F., G.S.C., A.N., N. J. \ \& M.P.M.D.\ performed the x-ray measurements. J.Z. \& J.F.M.\ synthesized the samples. G.F., D.M., Y.S., S.J., M.R.N., \& M.P.M.D. performed the technical validation of the data. All authors reviewed the manuscript. 

\section*{Competing interests}
The authors declare no competing interests.

\section*{Figures \& Tables}

\begin{thebibliography}{10}
\urlstyle{rm}
\expandafter\ifx\csname url\endcsname\relax
  \def\url#1{\texttt{#1}}\fi
\expandafter\ifx\csname urlprefix\endcsname\relax\def\urlprefix{URL }\fi
\expandafter\ifx\csname doiprefix\endcsname\relax\def\doiprefix{DOI: }\fi
\providecommand{\bibinfo}[2]{#2}
\providecommand{\eprint}[2][]{\url{#2}}

\bibitem{Anisimov1999electronic}
\bibinfo{author}{Anisimov, V.}, \bibinfo{author}{Bukhvalov, D.} \&
  \bibinfo{author}{Rice, T.}
\newblock \bibinfo{journal}{\bibinfo{title}{Electronic structure of possible
  nickelate analogs to the cuprates}}.
\newblock {\emph{\JournalTitle{Physical Review B}}}
  \textbf{\bibinfo{volume}{59}}, \bibinfo{pages}{7901} (\bibinfo{year}{1999}).

\bibitem{Norman2016materials}
\bibinfo{author}{Norman, M.~R.}
\newblock \bibinfo{journal}{\bibinfo{title}{Materials design for new
  superconductors}}.
\newblock {\emph{\JournalTitle{Reports on Progress in Physics}}}
  \textbf{\bibinfo{volume}{79}}, \bibinfo{pages}{074502}
  (\bibinfo{year}{2016}).

\bibitem{Adler2018correlated}
\bibinfo{author}{Adler, R.}, \bibinfo{author}{Kang, C.-J.},
  \bibinfo{author}{Yee, C.-H.} \& \bibinfo{author}{Kotliar, G.}
\newblock \bibinfo{journal}{\bibinfo{title}{Correlated materials design:
  prospects and challenges}}.
\newblock {\emph{\JournalTitle{Reports on Progress in Physics}}}
  \textbf{\bibinfo{volume}{82}}, \bibinfo{pages}{012504}
  (\bibinfo{year}{2019}).

\bibitem{Chaloupka2008orbital}
\bibinfo{author}{Chaloupka, J.} \& \bibinfo{author}{Khaliullin, G.}
\newblock \bibinfo{journal}{\bibinfo{title}{Orbital order and possible
  superconductivity in {${\mathrm{LaNiO}}_{3}/{\mathrm{LaMO}}_{3}$}
  superlattices}}.
\newblock {\emph{\JournalTitle{Phys. Rev. Lett.}}}
  \textbf{\bibinfo{volume}{100}}, \bibinfo{pages}{016404}
  (\bibinfo{year}{2008}).

\bibitem{Hansmann2009turninng}
\bibinfo{author}{Hansmann, P.} \emph{et~al.}
\newblock \bibinfo{journal}{\bibinfo{title}{Turning a nickelate fermi surface
  into a cupratelike one through heterostructuring}}.
\newblock {\emph{\JournalTitle{Phys. Rev. Lett.}}}
  \textbf{\bibinfo{volume}{103}}, \bibinfo{pages}{016401}
  (\bibinfo{year}{2009}).

\bibitem{Disa2015orbital}
\bibinfo{author}{Disa, A.~S.} \emph{et~al.}
\newblock \bibinfo{journal}{\bibinfo{title}{Orbital engineering in
  symmetry-breaking polar heterostructures}}.
\newblock {\emph{\JournalTitle{Phys. Rev. Lett.}}}
  \textbf{\bibinfo{volume}{114}}, \bibinfo{pages}{026801}
  (\bibinfo{year}{2015}).

\bibitem{Li2019superconductivity}
\bibinfo{author}{Li, D.} \emph{et~al.}
\newblock \bibinfo{journal}{\bibinfo{title}{Superconductivity in an
  infinite-layer nickelate}}.
\newblock {\emph{\JournalTitle{Nature}}} \textbf{\bibinfo{volume}{572}},
  \bibinfo{pages}{624--627} (\bibinfo{year}{2019}).

\bibitem{Osada2020superconducting}
\bibinfo{author}{Osada, M.} \emph{et~al.}
\newblock \bibinfo{journal}{\bibinfo{title}{A superconducting praseodymium
  nickelate with infinite layer structure}}.
\newblock {\emph{\JournalTitle{Nano Letters}}} \textbf{\bibinfo{volume}{20}},
  \bibinfo{pages}{5735--5740} (\bibinfo{year}{2020}).

\bibitem{Pan2021super}
\bibinfo{author}{Pan, G.~A.} \emph{et~al.}
\newblock \bibinfo{journal}{\bibinfo{title}{Superconductivity in a
  quintuple-layer square-planar nickelate}}.
\newblock {\emph{\JournalTitle{Nature Materials}}}
  \textbf{\bibinfo{volume}{21}}, \bibinfo{pages}{160--164}
  (\bibinfo{year}{2022}).

\bibitem{Zeng2022Superconductivity}
\bibinfo{author}{Zeng, S.} \emph{et~al.}
\newblock \bibinfo{journal}{\bibinfo{title}{Superconductivity in infinite-layer
  nickelate {La$_{1-x}$Ca$_x$NiO$_2$} thin films}}.
\newblock {\emph{\JournalTitle{Science Advances}}}
  \textbf{\bibinfo{volume}{8}}, \bibinfo{pages}{eabl9927}
  (\bibinfo{year}{2022}).

\bibitem{Hepting2021soft}
\bibinfo{author}{Hepting, M.}, \bibinfo{author}{Dean, M.~P.} \&
  \bibinfo{author}{Lee, W.-S.}
\newblock \bibinfo{journal}{\bibinfo{title}{Soft x-ray spectroscopy of
  low-valence nickelates}}.
\newblock {\emph{\JournalTitle{Frontiers in Physics}}}
  \textbf{\bibinfo{volume}{9}}, \bibinfo{pages}{777} (\bibinfo{year}{2021}).

\bibitem{Dean2015insights}
\bibinfo{author}{Dean, M.}
\newblock \bibinfo{journal}{\bibinfo{title}{Insights into the high temperature
  superconducting cuprates from resonant inelastic x-ray scattering}}.
\newblock {\emph{\JournalTitle{J. Magn. Magn. Mater.}}}
  \textbf{\bibinfo{volume}{376}}, \bibinfo{pages}{3--13}
  (\bibinfo{year}{2015}).

\bibitem{Fabbris2017doping}
\bibinfo{author}{Fabbris, G.} \emph{et~al.}
\newblock \bibinfo{journal}{\bibinfo{title}{Doping dependence of collective
  spin and orbital excitations in the spin-1 quantum antiferromagnet
  {${\mathrm{La}}_{2\ensuremath{-}x}{\mathrm{Sr}}_{x}{\mathrm{NiO}}_{4}$}
  observed by x rays}}.
\newblock {\emph{\JournalTitle{Physical Review Lett.}}}
  \textbf{\bibinfo{volume}{118}}, \bibinfo{pages}{156402}
  (\bibinfo{year}{2017}).

\bibitem{Zhang2017large}
\bibinfo{author}{Zhang, J.} \emph{et~al.}
\newblock \bibinfo{journal}{\bibinfo{title}{Large orbital polarization in a
  metallic square-planar nickelate}}.
\newblock {\emph{\JournalTitle{Nature Physics}}} \textbf{\bibinfo{volume}{13}},
  \bibinfo{pages}{864--869} (\bibinfo{year}{2017}).

\bibitem{Hepting2020electronic}
\bibinfo{author}{Hepting, M.} \emph{et~al.}
\newblock \bibinfo{journal}{\bibinfo{title}{Electronic structure of the parent
  compound of superconducting infinite-layer nickelates}}.
\newblock {\emph{\JournalTitle{Nature Materials}}}
  \textbf{\bibinfo{volume}{19}}, \bibinfo{pages}{381--385}
  (\bibinfo{year}{2020}).

\bibitem{Lin2021strong}
\bibinfo{author}{Lin, J.~Q.} \emph{et~al.}
\newblock \bibinfo{journal}{\bibinfo{title}{Strong superexchange in a
  ${d}^{9\ensuremath{-}\ensuremath{\delta}}$ nickelate revealed by resonant
  inelastic x-ray scattering}}.
\newblock {\emph{\JournalTitle{Physical Review Letters}}}
  \textbf{\bibinfo{volume}{126}}, \bibinfo{pages}{087001}
  (\bibinfo{year}{2021}).

\bibitem{Lu2021magnetic}
\bibinfo{author}{Lu, H.} \emph{et~al.}
\newblock \bibinfo{journal}{\bibinfo{title}{Magnetic excitations in
  infinite-layer nickelates}}.
\newblock {\emph{\JournalTitle{Science}}} \textbf{\bibinfo{volume}{373}},
  \bibinfo{pages}{213} (\bibinfo{year}{2021}).

\bibitem{Higashi2021core}
\bibinfo{author}{Higashi, K.}, \bibinfo{author}{Winder, M.},
  \bibinfo{author}{Kune\v{s}, J.} \& \bibinfo{author}{Hariki, A.}
\newblock \bibinfo{journal}{\bibinfo{title}{Core-level x-ray spectroscopy of
  infinite-layer nickelate: {$\mathrm{LDA}+\mathrm{DMFT}$} study}}.
\newblock {\emph{\JournalTitle{Physical Review X}}}
  \textbf{\bibinfo{volume}{11}}, \bibinfo{pages}{041009}
  (\bibinfo{year}{2021}).

\bibitem{Rossi2020orbital}
\bibinfo{author}{Rossi, M.} \emph{et~al.}
\newblock \bibinfo{journal}{\bibinfo{title}{Orbital and spin character of doped
  carriers in infinite-layer nickelates}}.
\newblock {\emph{\JournalTitle{Physical Review B}}}
  \textbf{\bibinfo{volume}{104}}, \bibinfo{pages}{L220505}
  (\bibinfo{year}{2021}).

\bibitem{Shen2022role}
\bibinfo{author}{Shen, Y.} \emph{et~al.}
\newblock \bibinfo{journal}{\bibinfo{title}{Role of oxygen states in the low
  valence nickelate {La$_{4}$Ni$_{3}$O$_{8}$}}}.
\newblock {\emph{\JournalTitle{Phys. Rev. X}}} \textbf{\bibinfo{volume}{12}},
  \bibinfo{pages}{011055} (\bibinfo{year}{2022}).

\bibitem{Norman2023orbital}
\bibinfo{author}{Norman, M.~R.} \emph{et~al.}
\newblock \bibinfo{title}{Orbital polarization, charge transfer, and
  fluorescence in reduced valence nickelates} (\bibinfo{year}{2023}).
\newblock \bibinfo{note}{ArXiv:2302.09003}.

\bibitem{Shen2023orbital}
\bibinfo{author}{Shen, Y.} \emph{et~al.}
\newblock \bibinfo{journal}{\bibinfo{title}{Electronic character of charge
  order in square-planar low-valence nickelates}}.
\newblock {\emph{\JournalTitle{Phys. Rev. X}}} \textbf{\bibinfo{volume}{13}},
  \bibinfo{pages}{011021} (\bibinfo{year}{2023}).

\bibitem{Wang2019EDRIXS}
\bibinfo{author}{Wang, Y.}, \bibinfo{author}{Fabbris, G.},
  \bibinfo{author}{Dean, M.} \& \bibinfo{author}{Kotliar, G.}
\newblock \bibinfo{journal}{\bibinfo{title}{{EDRIXS}: An open source toolkit
  for simulating spectra of resonant inelastic x-ray scattering}}.
\newblock {\emph{\JournalTitle{Computer Physics Communications}}}
  \textbf{\bibinfo{volume}{243}}, \bibinfo{pages}{151--165}
  (\bibinfo{year}{2019}).

\bibitem{Krieger2021Charge}
\bibinfo{author}{Krieger, G.} \emph{et~al.}
\newblock \bibinfo{journal}{\bibinfo{title}{Charge and spin order dichotomy in
  {${\mathrm{NdNiO}}_{2}$} driven by the capping layer}}.
\newblock {\emph{\JournalTitle{Physical Review Letters}}}
  \textbf{\bibinfo{volume}{129}}, \bibinfo{pages}{027002}
  (\bibinfo{year}{2022}).

\bibitem{Zhang2016stacked}
\bibinfo{author}{Zhang, J.} \emph{et~al.}
\newblock \bibinfo{journal}{\bibinfo{title}{Stacked charge stripes in the
  quasi-{2D} trilayer nickelate {La$_4$Ni$_3$O$_8$}}}.
\newblock {\emph{\JournalTitle{Proceedings of the National Academy of
  Sciences}}} \textbf{\bibinfo{volume}{113}}, \bibinfo{pages}{8945--8950}
  (\bibinfo{year}{2016}).

\bibitem{Chiuzbaian2014}
\bibinfo{author}{Chiuzbăian, S.~G.} \emph{et~al.}
\newblock \bibinfo{journal}{\bibinfo{title}{Design and performance of {AERHA},
  a high acceptance high resolution soft x-ray spectrometer.}}
\newblock {\emph{\JournalTitle{Rev. Sci. Instrum.}}}
  \textbf{\bibinfo{volume}{85}}, \bibinfo{pages}{043108}
  (\bibinfo{year}{2014}).

\bibitem{Zenodo}
\bibinfo{note}{Zenodo Data Repository
  \url{https://doi.org/10.5281/zenodo.7582152} (2023).}

\bibitem{harris2020array}
\bibinfo{author}{Harris, C.~R.} \emph{et~al.}
\newblock \bibinfo{journal}{\bibinfo{title}{Array programming with {NumPy}}}.
\newblock {\emph{\JournalTitle{Nature}}} \textbf{\bibinfo{volume}{585}},
  \bibinfo{pages}{357--362} (\bibinfo{year}{2020}).

\bibitem{Hunter2007matplotlib}
\bibinfo{author}{Hunter, J.~D.}
\newblock \bibinfo{journal}{\bibinfo{title}{Matplotlib: A 2d graphics
  environment}}.
\newblock {\emph{\JournalTitle{Computing in Science \& Engineering}}}
  \textbf{\bibinfo{volume}{9}}, \bibinfo{pages}{90--95} (\bibinfo{year}{2007}).

\bibitem{mckinney2010data}
\bibinfo{author}{McKinney, W.} \emph{et~al.}
\newblock \bibinfo{title}{Data structures for statistical computing in python}.
\newblock In \emph{\bibinfo{booktitle}{Proceedings of the 9th Python in Science
  Conference}}, vol. \bibinfo{volume}{445}, \bibinfo{pages}{51--56}
  (\bibinfo{organization}{Austin, TX}, \bibinfo{year}{2010}).

\end{thebibliography}
\end{document}